\documentclass{elsart1p}
\usepackage[ruled]{algorithm}
\usepackage[noend]{algorithmic}
\usepackage{amsmath}
\usepackage{amssymb}
\usepackage{graphicx}
\input{Qcircuit}

\def\C{\mathbb{C}}
\def\Z{\mathbb{Z}}

\def\w{\omega}

\begin{document}

\begin{frontmatter}

  \title{Obtaining the Quantum Fourier Transform from the Classical
    FFT with QR Decomposition}

  \author[LNCC]{F.L. Marquezino}, \ead{franklin@lncc.br}
  \author[LNCC]{R. Portugal}, \ead{portugal@lncc.br}
  \author[UDESC]{F.D. Sasse} \ead{fsasse@joinville.udesc.br}

  \address[LNCC]{Laborat\'orio Nacional de Computa\c{c}\~ao
    Cient\'{\i}fica\\Rua Get\'ulio Vargas, 333, Petr\'opolis,
    25651-075, RJ, Brazil} \address[UDESC]{Universidade do Estado de
    Santa Catarina\\Dept. of Mathematics, CCT, Joinville, 89223-100,
    SC, Brazil}

\begin{abstract}
  We present the detailed process of converting the classical Fourier
  Transform algorithm into the quantum one by using QR
  decomposition. This provides an example of a technique for building
  quantum algorithms using classical ones. The Quantum Fourier
  Transform is one of the most important quantum subroutines known at
  present, used in most algorithms that have exponential speed up
  compared to the classical ones. We briefly review Fast Fourier
  Transform and then make explicit all the steps that led to the
  quantum formulation of the algorithm, generalizing Coppersmith's
  work.
\end{abstract}
\end{frontmatter}

\section{Introduction}

For many years the spectral analysis of sampled data over a finite
range, referred as the Discrete Fourier Transform (DFT) was performed
directly on computers, using $O(N^2)$ operations, where $N$ is the
number of data points.  A great milestone in Fourier analysis was the
paper published in 1965 by Cooley and Tukey~\cite{CT65}, where they
described the so-called Fast Fourier Transform (FFT) algorithm, which
can compute the DFT with only $O(N \log N)$ operations.  In the next
year, Rudnick presented a computer program which also required $O(N
\log N)$ operations~\cite{Rud66}, being inspired by an earlier method
due to Danielson and Lanczos~\cite{DL42}.

Although the FFT algorithm may seem reasonably fast for classical
Computer Science, it turns out that its quantum version can provide
exponentially faster algorithms for some problems. In 1994,
Shor~\cite{Sho94a} developed a quantum version of the Fourier
Transform when the prime factors of $N$ are not large (smaller than
$\log N$). Motivated by Shor's work, Coppersmith~\cite{Cop94a}
developed the quantum version of the FFT when $N$ is a power of two,
which required only $O(\log^2 N)$ operations. In the same year
Cleve~\cite{Cle94}, using a recursive approach, has also shown how to
implement the Fourier Transform in quantum computers.\footnote{Nielsen
  and Chuang~\cite{NC00a} also mention an unpublished work on this
  subject by David Deutsch.} This is, with no doubt, the most
important quantum sub-routine designed so far. Most quantum algorithms
with exponential improvement when compared to the classical
counterparts use the Quantum Fourier Transform (QFT) as an essential
part. These algorithms solve instances of the Hidden Subgroup Problem
(HSP), which has several important applications. A good survey on the
HSP can be found in~\cite{Lom04b}.

The quantum algorithm for an approximate QFT was developed by
Coppersmith~\cite{Cop94a}, with complexity $O(m \log N)$, where $1
\leq m \leq \log N$ is the parameter that defines the degree of
approximation, usually taken as $O(\log \log N)$. Indeed, this
approximate algorithm is the one with practical applications. The
exact transform yields a level of accuracy that, in most of the cases,
cannot even be detected by the measuring device. Barenco \textit{et
  al.}~\cite{BEST96a} showed that, in the presence of decoherence,
i.e., in realistic situations, the approximate transform may achieve
better results than the exact one. They also showed that the accuracy
of the exact QFT can be achieved by applying the approximate QFT
repeatedly $O\left(\frac{\log^3{N}}{m^3}\right)$ times.

In this paper we build the QFT algorithm from the classical FFT,
generalizing Coppersmith's work. In order to accomplish this
generalization, we use the QR decomposition. This technique may
provide insight for the development of new quantum algorithms.

In Sec.~2 we review the definition of DFT and the classical FFT
algorithm. We also fix the notation that will be used in the
generalization of the quantum algorithm. In Sec.~3 we perform the
decomposition of the matrix form of FFT algorithm. QR decomposition
plays an essential role in this task. The resulting unitary matrices
can be interpreted as quantum logic gates. In Sec.~4 we summarize the
results obtained in the previous sections and show how to build the
circuit for the exact QFT.

\section{The Discrete Fourier Transform}\label{sec:dft}

In this section we address the DFT in a form that will be useful later
on. It is important to establish some notations that will be used
throughout this paper.  Let $n$ be a positive integer, and let $a,c$
be $n$-bit integers. The binary representations of $a$ and $c$ will be
denoted, respectively, by $$(a_{n-1} a_{n-2} \ldots a_0)_2 \mbox{ and
} (c_{n-1} c_{n-2} \ldots c_0)_2,$$ such that,
$$a = \sum_{j=0}^{n-1}{a_j 2^j} \mbox{ and } c = \sum_{j=0}^{n-1}{c_j 2^j}.$$ 

Binary representations of numbers may have the subscript $(\cdot)_2$
omitted when context is clear. All logarithms are base two. Let $X$
and $Y$ be arrays\footnote{Some readers may prefer to see $X$ and $Y$
  as functions, respectively $X:\Z_N \rightarrow \C$, and $Y:\Z_N
  \rightarrow \C$, where $N=2^n$.} of $N=2^n$ complex numbers.  The
integers $a$ and $c$ defined above will be useful for indexing $X$ and
$Y$. The notation used in this paper for that indexing is $X_a$.  Let
$\omega \equiv \omega_{N} = \exp(2\pi i/N),$ be the $N$-root of unity.

Finally, we may define the DFT.

\begin{defn}[DFT]
  The Discrete Fourier Transform takes as input a complex array $X$,
  and converts it into a complex array $Y$, according to the
  expression
  \begin{equation}
    Y_c = \dfrac{1}{\sqrt{N}} \sum_{a=0}^{N-1}{X_a \omega^{ac}} .
    \label{eq:dft}
  \end{equation}
  \label{def:dft}
\end{defn}

This equation may be rewritten by making explicit the binary
representation of $a$ and $c$. We concentrate only on the sub-term
$\omega^{ac}$.

\begin{equation}
  \omega^{ac} = \exp\left(\dfrac{2\pi i}{N} \sum_{0 \leq j,k \leq n-1}{a_j c_k 2^{j+k}} \right).
  \label{eq:dft2}
\end{equation}
Using $\omega^{(2^n)} = 1,$ we can state that
\begin{equation}
  \omega^{(2^{j+k})} = \omega^{{(2^n)}^{2^{j+k-n}}} = 1,
\end{equation}
if $j+k \geq n.$ Therefore, it is possible to eliminate terms from
Eq.~(\ref{eq:dft2}) when $j+k\geq n$, leading us to an equivalent
definition of the DFT,
\begin{equation}
  Y_c = \frac{1}{\sqrt{N}} \sum_{0 \leq a \leq N-1}{X_a \exp\Biggl(\frac{2\pi i}{N} 
    \sum_{\substack{0 \leq j,k \leq n-1\\j+k \leq n-1}}{a_j c_k 2^{j+k}} \Biggr)}.
  \label{eq:dftCopp}
\end{equation}

Coppersmith~\cite{Cop94a} observed that we may define a more general
transform by changing the range of $j+k$ in Eq.~\eqref{eq:dftCopp}. If
we take $n-m\leq j+k \leq n-1$, where $m$ is a parameter such that
$1\leq m \leq n$, we obtain the definition of the Approximate Fourier
Transform. When $m=n$, we recover the definition of the exact
DFT. When $m=1$, we achieve the a slighty modified definition of the
well known Hadamard transform---with a different indexing of the
elements in the array.  The argument in the exponent of this new
transform differs from that in the exponent of DFT just by
\begin{equation}
  i\epsilon = \dfrac{2\pi i}{N} \sum_{\substack{0 \leq j,k \leq n-1\\j+k<n-m}}{a_j c_k 2^{j+k}}.
  \label{eq:dif}
\end{equation}
It is not difficult to show that the magnitude $|\epsilon|$ of that
error in Eq.~(\ref{eq:dif}) is bounded by $2\pi n 2^{-m}$.  Both the
approximate and the exact Fourier Transforms may be expressed by a
matrix operation over the array $X$.  The entries of the corresponding
matrices only differ by a multiplicative factor $\exp{(i\epsilon)},$
where $|\epsilon| \leq 2\pi n 2^{-m}.$ Therefore, the approximate
transform is able to provide results very close to the exact Fourier
Transform, even if the parameter $m$ is not so close to $n$, because
the error decreases exponentially as $m$ increases.  The minimum $m$
required to ensure an error not greater than $|\epsilon_{max}|$ is
given by
\begin{equation}
  m = \log{\frac{2\pi}{|\epsilon_{max}|}} + \log \log{N}.
\end{equation}
Thus, the Approximate Fourier Transform provides results with a fixed
tolerance $|\epsilon_{max}|$ by using a parameter $m$ that barely
increases with the size of the input.

We now briefly review the classical FFT~\cite{Knu81a,NRC92} via
Danielson Lanczos lemma.  Danielson Lanczos lemma~\cite{DL42} allows
the development of a recursive divide-and-conquer scheme to calculate
the DFT. According to this lemma, Eq.~(\ref{eq:dft}) can be split into
two parts. These parts of length $N/2$ can be calculated by employing
again Danielson Lanczos lemma, leading to the calculation of four new
Fourier Transforms, of length $N/4$ each. This process repeats itself
$n$ times until hitting arrays of length one, for which the Fourier
Transforms are trivially found---in Eq.~(\ref{eq:dft}), when $N=1$, we
have $Y=X$. This is the idea behind Algorithm~\ref{alg:copp}. Here, we
denote by $X^{(s)}$ the state of a vector $X$ in a step~$s$ of the
computation.

\begin{algorithm}[h!]
  \caption{Classical FFT}
  \label{alg:copp}
  \begin{algorithmic}[1]
    \REQUIRE This algorithm receives as input a vector $X \in
    \C^{2^n}$.  \ENSURE The output is a vector $Y \in \C^{2^n}$ which
    is the DFT of vector $X$.  \FORALL[Initialization, step $n$]{$a$
      such that $0 \leq a \leq 2^n-1$} \STATE let $X^{(n)}_{(a_{n-1}
      a_{n-2} \ldots a_0)_2} \leftarrow X_{(a_{n-1} a_{n-2} \ldots
      a_0)_2}.$
    \ENDFOR
    \FOR[Step $s$]{$s$ from $n-1$ to $0$, downward} \FORALL{$0 \leq
      b_{n-1}, \ldots, b_s, a_{s-1}, \ldots, a_0 \leq 1$} \STATE
    \begin{eqnarray}
      X^{(s)}_{(b_{n-1} \ldots b_s a_{s-1} \ldots a_0)_2} \leftarrow
      \frac{1}{\sqrt{2}} X^{(s+1)}_{(b_{n-1} \ldots b_{s+1} 0 a_{s-1} \ldots a_0)_2} +\nonumber\\
      + \frac{1}{\sqrt{2}} \w^{(b_s b_{s+1} \ldots b_{n-1} 0 \ldots 0)_2}
      X^{(s+1)}_{(b_{n-1} \ldots b_{s+1} 1 a_{s-1} \ldots a_0)_2}.\label{eq:alg}
    \end{eqnarray}
    \ENDFOR
    \ENDFOR
    \FORALL[Re-ordering]{$b$ such that $0 \leq b \leq 2^n-1$} \STATE
    set $Y_{(b_{n-1} b_{n-2} \ldots b_0)_2} \leftarrow X^{(0)}_{(b_0
      b_1 \ldots b_{n-1})_2}.$
    \ENDFOR
  \end{algorithmic}
\end{algorithm}

The complexity of the FFT can be easily computed from the above
considerations. If we denote by $T_{2^n}$ the approximate number of
steps of a DFT on an array of length $2^n$, we may write $T_{2^n} =
2T_{2^{n-1}}+2^n$, for $n\geq 1$, with $T_1=0$.
Then, $T_{2^n} = n2^n$, which means that the FFT algorithms have
complexity $O(n 2^n)$ or, equivalently, $O(N \log N)$.

\section{Obtaining QFT from FFT}\label{sec:ctoquantum}

If we treat $X^{(s)}$ as column vectors, it becomes clear that
Algorithm~\ref{alg:copp} may be expressed as a matrix operation, such
that
\begin{equation}
  X^{(s)}_j = \sum_{0 \leq k \leq N-1}{P^{(s)}_{jk} X^{(s+1)}_k},
  \label{eq:alg2matrix}
\end{equation}
where $P^{(s)}$ is a $N \times N$ matrix for $0 \leq s \leq n-1$. The
indices of vectors, as well as the rows and columns of the matrices
will be numbered from $0$ to $N-1$ in this paper.

By observing Algorithm~\ref{alg:copp}, we see in Eq.~(\ref{eq:alg})
that each row of $X^{(s)}$ depends only on two rows of
$X^{(s+1)}$. Hence, each row of the matrices $P^{(s)}$ have only two
nonzero entries. They are located on columns $k=j-j_s 2^s$ and
$k=j+(1-j_s) 2^s$, where $j_s$ is the $(s+1)$-th bit of $j$ (counting
from the least to the most significant bit).  Thus, each nonzero entry
of matrices $P^{(s)}$ will be placed only in the main diagonal, or in
a subdiagonal, depending of the value of the bit $j_s$ of $j$.

When $k=j$, we have the entries of the main diagonal of the
matrices. If $j_s=0$, then the first term on the right hand side of
Eq.~(\ref{eq:alg}) shows us that these entries are
$\frac{1}{\sqrt{2}}$.  If $j_s=1$, then the second term on the right
hand side of Eq.~(\ref{eq:alg}) shows us that the entries are
$\frac{1}{\sqrt{2}}\w^{(j_s j_{s+1} \ldots j_{n-1} 0 \ldots 0)_2}$.

When $k=j-2^s$, we have the entries of the lower subdiagonal of the
matrices. If $j_s=0$, then the entries are $0$, according to
Eq.~(\ref{eq:alg}). If $j_s=1$, then the first term on the right hand
side of Eq.~(\ref{eq:alg}) shows us that the entries are
$\frac{1}{\sqrt{2}}$.

When $k=j+2^s$, we have the entries of the upper subdiagonal of the
matrices. If $j_s=0$, then the second term on the right hand side of
Eq.~(\ref{eq:alg}) shows us that the entries are
$\frac{1}{\sqrt{2}}\w^{(j_s j_{s+1} \ldots j_{n-1} 0 \ldots 0)_2}$.
If $j_s=1$, then the entries are $0$, according to Eq.~(\ref{eq:alg}).
Before we express these ideas mathematically let us introduce
additional notation.

If we define a binary fraction as
\begin{equation}
  0.j \equiv 0.j_0j_1 \cdots j_{n-1} = \sum_{0 \leq t \leq n-1}{\frac{j_t}{2^{t+1}}},
  \label{eq:jbinaryfraction}
\end{equation}
we may rewrite $\w^{(j_s j_{s+1} \ldots j_{n-1} 0 \ldots 0)_2}$ as
$\w^{(0.j) 2^{n+s} }$.

We note also that
\begin{equation}
  j_s = \dfrac{1-(-1)}{2}^{{}^{\left\lfloor \dfrac{j}{2^s} \right\rfloor}}.
  \label{eq:sthbitofj}
\end{equation}
The generic matrices $P^{(s)}$ can now be written as,
\begin{equation}
  P^{(s)}_{jk} = \frac{1}{\sqrt{2}}
  \begin{cases}
    \omega^{j_s (0.j) 2^{n+s}}, & \mbox{if } k=j\\
    \dfrac{1-(-1)}{2}^{\left\lfloor \frac{j}{2^s} \right \rfloor}, &\mbox{if }k=j-2^s\\
    \dfrac{1+(-1)}{2}^{\left\lfloor \frac{j}{2^s} \right\rfloor} \omega^{(0.j) 2^{n+s}}, & \mbox{if } k=j+2^s\\
    0, & \mbox{otherwise,}
  \end{cases}
  \label{eq:PGenerica}
\end{equation}
which represent the operations performed by the steps of
Algorithm~\ref{alg:copp}.

\begin{prop}
  The matrices $P^{(s)}$ are unitary.
\end{prop}
\begin{pf}
  We may solve
  \begin{equation}
    \left(P^{(s)}P^{(s)^\dagger}\right)_{jk} = \sum_{0 \leq l \leq N-1}{P^{(s)}_{jl}P^{(s)^\ast}_{kl}}.
  \end{equation}

  When $k=j$, we have
  \begin{eqnarray}
    \left(P^{(s)}P^{(s)^\dagger}\right)_{jj} &=& P^{(s)}_{jj}P^{(s)^\ast}_{jj} + P^{(s)}_{j,j-2^s}P^{(s)^\ast}_{j,j-2^s} +
    P^{(s)}_{j,j+2^s}P^{(s)^\ast}_{j,j+2^s}\nonumber\\
    &=&
    \dfrac{1}{2} +
    \dfrac{1}{2} \left( \dfrac{1-(-1)}{2}^{\left\lfloor \frac{j}{2^s} \right\rfloor} \right)^2 +
    \dfrac{1}{2} \left( \dfrac{1+(-1)}{2}^{\left\lfloor \frac{j}{2^s}\right\rfloor} \right)^2
    \nonumber\\
    &=& 1.
  \end{eqnarray}

  When $k=j-2^s$ or $k=j+2^s$ we may perform analogous calculation and
  obtain $\left(P^{(s)}P^{(s)^\dagger}\right)_{jk} = 0$.  Therefore,
  $\left(P^{(s)}P^{(s)^\dagger}\right)_{jk} = \delta_{jk}.$ $\Box$
\end{pf}

Since the matrices $P^{(s)}$ are unitary, they could in principle be
implemented on a quantum computer.  It is important, when developing a
quantum algorithm, to express the unitary operators in terms of
universal quantum gates, that is, controlled-NOTs and gates acting on
single qubits.

In order to find the universal gates for the quantum version of the
FFT algorithm, several decompositions may be applied to matrices
$P^{(s)}$. One method that may be insightful in this process is the QR
decomposition~\cite{GVL96,NRC92}, which factors a generic matrix into
the product of a unitary matrix $M$---orthogonal, if the matrix to be
decomposed is real---and an upper triangular matrix $N$.  In the case
of matrices $P^{(s)}$ we may apply a slightly modified version of QR
decomposition which yields orthogonal and diagonal matrices as
factors. This can be done because matrices $P^{(s)}$ have the
following property: their columns are either real or multiple of real
columns.  There are at least three well known methods for computing
the QR decomposition: Householder reflections, Givens rotations and
Gram-Schmidt decomposition.  The last one is particularly interesting
in this case because it takes advantage of the property mentioned
above.

Observe that any column of matrices $P^{(s)}$ may be obtained by
fixing a value of $k$ in Eq.~(\ref{eq:PGenerica}) and running $j$ from
$0$ to $N-1$. We note that the columns of matrices $P^{(s)}$ are
already orthonormal. Multiplying each column $k$ by
\begin{equation}
  \alpha^{(s)}_k \equiv (-1)^{\left\lfloor \frac{k}{2^s} \right\rfloor} \omega^{2^n - k_s (0.k) 2^{n+s}},
\end{equation}
for $k=j$, $k=j-2^s$ and $k=j+2^s$, and then multiplying the
corresponding cases in Eq.~(\ref{eq:PGenerica}) by $\alpha^{(s)}_k$,
we obtain the orthogonal matrices
\begin{equation}
  M^{(s)}_{jk} =
  \dfrac{1}{\sqrt{2}}
  \begin{cases}
    (-1)^{\left\lfloor \frac{j}{2^s} \right\rfloor}, & \mbox{if } k=j\\
    \dfrac{1-(-1)}{2}^{\left\lfloor \frac{j}{2^s} \right \rfloor}, & \mbox{if } k=j-2^s\\
    \dfrac{1+(-1)}{2}^{\left\lfloor \frac{j}{2^s} \right \rfloor}, & \mbox{if } k=j+2^s\\
    0, & \mbox{otherwise.}
  \end{cases}
  \label{eq:QGenerica}
\end{equation}
The upper triangular matrices are
\begin{equation}
  N^{(s)}_{jk} =
  \begin{cases}
    (-1)^{\left\lfloor \frac{j}{2^s} \right\rfloor} \omega^{j_s(0.j)2^{n+s}}, & \mbox{if } k=j\\
    0, & \mbox{otherwise.}
  \end{cases}
  \label{eq:RGenerica}
\end{equation}

Now we confirm the decomposition.
\begin{prop}
  For any step $s$ and for any number of bits $n$ we have
  \begin{equation}
    P^{(s)} = M^{(s)} N^{(s)},
  \end{equation}
  where the matrices $M^{(s)}$ and $N^{(s)}$ are given by
  Eqs.~(\ref{eq:QGenerica}) and~(\ref{eq:RGenerica}), respectively.
\end{prop}
\begin{pf}
  Since matrices $N^{(s)}$ are diagonal we have
  \begin{equation}
    \left(M^{(s)} N^{(s)}\right)_{jk} = M^{(s)}_{jk} N^{(s)}_{kk}.
  \end{equation}

  When $k=j$,
  \begin{eqnarray}
    M^{(s)}_{jj} N^{(s)}_{jj}
    &=&
    \frac{(-1)}{\sqrt{2}}^{\left\lfloor \frac{j}{2^s} \right\rfloor}
    (-1)^{\left\lfloor \frac{j}{2^s} \right\rfloor}
    \w^{j_s (0.j) 2^{n+s}}\nonumber\\
    &=& P_{jj}^{(s)}.
  \end{eqnarray}

  When $k=j-2^s$ or $k=j+2^s$ we may perform analogous calculations,
  and easily obtain $\left(M^{(s)} N^{(s)}\right)_{jk} =
  P^{(s)}_{jk}.$ $\Box$
\end{pf}

Let us now analyze the structure of the matrices $M^{(s)}$. Starting
with $M^{(0)}$, we note that
\begin{equation}
  M_{jk}^{(0)} = \frac{1}{\sqrt{2}}
  \begin{cases}
    (-1)^j, & \mbox{if } j=k\\
    \dfrac{1-(-1)}{2}^j, & \mbox{if } k=j-1\\
    \dfrac{1+(-1)}{2}^j, & \mbox{if } k=j+1\\
    0, &\mbox{otherwise.}
  \end{cases}
\end{equation}
It is easy to check that
\begin{equation}
  M_{2^n \times 2^n}^{(0)} = I^{\otimes (n-1)} \otimes H,
  \label{eq:IIH}
\end{equation}
where $I$ is the $2 \times 2$ identity matrix, and
$H=\frac{1}{\sqrt{2}}\left( \begin{array}{cc}1&1\\1&-1\end{array}
\right)$ is the Hadamard matrix.

Now, we prove that $M_{2^n \times 2^n}^{(s)} = M_{2^{n-1} \times
  2^{n-1}}^{(s-1)} \otimes I$. We use the following formula for a
generic matrix $A$:
\begin{equation}
  (A \otimes I)_{jk} =
  \begin{cases}
    A_{\left\lfloor \dfrac{j}{2} \right\rfloor \left\lfloor \dfrac{k}{2} \right\rfloor}, & \mbox{if } k=j \pmod 2\\
    0, & \mbox{otherwise,}
  \end{cases}
  \label{eq:FormulaDoRenato}
\end{equation}
which can be obtained by analyzing the entries of $A \otimes I$.
Replacing $A$ by $M_{2^{n-1} \times 2^{n-1}}^{(s-1)}$ in
Eq.~(\ref{eq:FormulaDoRenato}) and simplifying the result, we get the
right-hand side of Eq.~(\ref{eq:QGenerica}). Thus
\begin{equation}
  M_{2^n \times 2^n}^{(s)} = M_{2^{n-1} \times 2^{n-1}}^{(s-1)} \otimes I.
  \label{eq:QI}
\end{equation}

\begin{prop}\label{pro:decomp}
  The matrices $M^{(s)}$ may be decomposed in tensorial products
  involving only $2\times 2$ identities and a Hadamard matrix, such
  that
  \begin{equation}
    M^{(s)} = I^{\otimes (n-s-1)}
    \otimes H \otimes I^{\otimes s}.
    \label{eq:QFat}
  \end{equation}
\end{prop}
\begin{pf}
  Using (\ref{eq:QI}) recursively $s$~times and replacing $n$ by $n-s$
  in Eq.~(\ref{eq:IIH}) we get Eq.~(\ref{eq:QFat}).  $\Box$
\end{pf}

Let us now analyze the structure of the matrices $N^{(s)}$. They are
diagonal and the entries are one or $\pm \w^c$, $0 \leq c < N$. The
first attempt in decomposing them is to write $N^{(s)}$ for a fixed
$s$ as a product of diagonal matrices, the entries of which are one or
$\pm \w^c$ for a fixed $c$. We rewrite Eq.~(\ref{eq:RGenerica}) using
Eqs.~(\ref{eq:jbinaryfraction}) and~(\ref{eq:sthbitofj}) to obtain the
form
\begin{equation}
  N_{jk}^{(s)} =
  \begin{cases}
    (-1)^{j_s}\prod_{t=0}^{n-1}\w^{j_s j_t 2^{n+s-t-1}} ,& \mbox{if } k=j,\\
    0, & \mbox{otherwise.}
  \end{cases}
  \label{eq:Ns}
\end{equation}
Since $n$ and $s$ are fixed, the factors we are looking for are
obtained by fixing $t$. So, for given $n$ and $s$, we define the
matrix
\begin{equation}
  R^{(s,t,u)} =
  \begin{cases}
    \w^{j_s j_t 2^{n-u}}, & \mbox{if } k=j\\
    0 , & \mbox{otherwise},
  \end{cases}
  \label{eq:RProd}
\end{equation}
where $j_s$ and $j_t$ are given by Eq.~(\ref{eq:sthbitofj}). Since
matrices $R^{(s,t,u)}$ are diagonal for arbitrary $s,t$ and $u$, they
commute. We can now state the following proposition.

\begin{prop}\label{pro:Ndecomp}
  The matrices $N^{(s)}$ may be written as
  \begin{equation}
    N^{(s)} = \prod_{t=s+1}^{n-1}{R^{(s,t,u)}},
    \label{eq:NProd}
  \end{equation}
  with $u=t-s+1$, when $s<n-1$. When $s=n-1$, $N^{(n-1)}=I$.
\end{prop}
\begin{pf}
  Using $(-1)^{j_s} = \w^{j_s j_s 2^{n-1}}$, we see that
  \begin{equation}
    (-1)^{j_s} \prod_{t=0}^{s}{\w^{j_s j_t 2^{n+s-t-1}}} = 1.
  \end{equation}
  Then, using this result in Eq.~(\ref{eq:Ns}) we get
  Eq.~(\ref{eq:NProd}) when $s<n-1$.  Replacing $s$ by $n-1$ in
  Eq.~(\ref{eq:Ns}) we get $N^{(n-1)} = I$.  $\Box$
\end{pf}

The structure of the matrices $R^{(s,t,u)}$ is the following: they are
diagonal; the entries are one when either $j_s$ or $j_t$ is equal to
zero; the entries are $\w^{2^{n-u}} = \exp{\left(\frac{2\pi
      i}{2^u}\right)}$ when $j_s$ and $j_t$ are simultaneously equal
to one. Therefore, the matrix $R^{(s,t,u)}$ is a controlled operation
\begin{equation}
  R^{(u)} \equiv \left( 
    \begin{array}{cc}
      1 & 0 \\ 
      0 & \exp\left( \frac{2\pi i}{2^u}\right)
    \end{array}
  \right),
  \label{eq:Ru}
\end{equation}
with control on qubit $s$ and target on qubit $t$ (or vice-versa, in
this case). This is a generalization of control gates acting on two
qubits in the presence of more qubits.

Summarizing, the QFT can be expressed as
\begin{equation}
  F_{2^n} = A^{(n)}\prod_{s=0}^{n-1}{M^{(s)}N^{(s)}},
  \label{eq:F-algebraic}
\end{equation}
where the matrices $M^{(s)}$ and $N^{(s)}$ are given by
Eqs.~(\ref{eq:QFat}) and~(\ref{eq:NProd}) respectively. $A^{(n)}$ is a
$2^n \times 2^n$ matrix implementing the final swaps of the algorithm.

\section{Building the QFT from the Proposed
  Decomposition}\label{sec:qftExact}

Based on the last section we may finally derive a quantum algorithm to
compute the DFT. In the beginning of classical FFT, we have a
collection of complex numbers $X^{(n)}_{(a_{n-1} a_{n-2} \cdots
  a_0)}.$ These values now correspond to the quantum state
\begin{equation}
  \ket{\psi_n} = \sum_{0 \leq a_{n-1}, a_{n-2}, \ldots , a_0 \leq 1}
  {X^{(n)}_{(a_{n-1} a_{n-2} \cdots a_0)} \ket{a_{n-1} a_{n-2} \ldots a_0}}.
\end{equation}
This preparation of the quantum system corresponds to the
initialization of the algorithm. Once the state has been prepared, we
should apply the matrices $P^{(s)}$ given by
Eq.~({\ref{eq:PGenerica}}) in the following order:
\begin{equation}
  \ket{\psi_0} = P^{(0)} P^{(1)} \ldots P^{(n-1)} \ket{\psi_n}.
\end{equation}
The state $\ket{\psi_0}$ is
\begin{equation}
  \ket{\psi_0} = \sum_{0 \leq c_{n-1} \cdots c_{0} \leq 1}{X^{(0)}_{(c_0 \cdots c_{n-1})} \ket{c_{n-1} \cdots c_0}},
\end{equation}
with the coefficients labelled in the inverse order.

\begin{algorithm}
  \caption{QFT}
  \label{alg:qft}
  \begin{algorithmic}[1]
    \REQUIRE This algorithm must receive as input a vector $X \in
    \C^{2^n}$.  \ENSURE The output is a quantum state $\ket{\psi}$
    whose amplitudes correspond to the elements of $Y \in \C^{2^n}$,
    given by the DFT of $X$.   
    \STATE \COMMENT{Initialization, step $n$}
    \STATE prepare the state of the $n$-qubit quantum register as 
       $$\ket{\psi_n} = \sum_{k=0}^{N-1}{X_k \ket{k}}.$$ 
       \FOR[Step $s$]{$s$ from $n-1$ to $0$, downward} \FOR{$t$ from
         $n-1$ to $s+1$, downward} \STATE apply unitary operation
       $R^{(s,t,t-s+1)}$
       \ENDFOR
       \STATE apply a Hadamard gate only on qubit $s$.
       \ENDFOR
       \STATE \FOR[Re-ordering]{$t$ from $0$ to $\lfloor n/2 \rfloor -
         1$} \STATE swap qubits $t$ and $n-t-1$.
       \ENDFOR
     \end{algorithmic}
   \end{algorithm}

   Although the matrices $P^{(s)}$ are, in general, too complex to be
   directly realized in a physical experiment, it was shown that each
   of them can be decomposed into simpler matrices $M^{(s)}$ and
   $N^{(s)}$, which in turn may be decomposed into gates acting on one
   or two qubits.

   In each step $s$, we must first apply the matrix $N^{(s)}$. In step
   $n-1$ we have $N^{(n-1)} = I$. Eq.~(\ref{eq:NProd}) shows us that
   each matrix $N^{(s)}$ is a product of other simpler
   matrices. Hence, we must apply the logical gate
   $R^{(s,t,t-s+1)}$---gate $R^{(t-s+1)}$ with control on qubit $s$
   and target on qubit $t$---for each $t$ starting from $t=n-1$ and
   going downward until $t=s+1$.  Then, we apply the matrix $M^{(s)}$,
   which corresponds to a Hadamard gate acting only on qubit
   $s$. After running $s$ from $n-1$ to $0$, we must apply swaps to
   correct the order of the output.  We compiled these steps in
   Algorithm~\ref{alg:qft}.

\begin{figure}
  \centering
  \begin{footnotesize}
$$
\Qcircuit @C=.5em @R=1.2em{
  & & & N^{(3)} & M^{(3)} & N^{(2)} & M^{(2)} & \mbox{  }N^{(1)} & & M^{(1)} & & N^{(0)} & & M^{(0)}\\
  &\lstick{3} &
  \qw & \gate{\mbox{ }I\mbox{ }} & \gate{\mbox{ }H\mbox{ }} & \gate{R^{(2)}} & \push{\rule{0em}{1.4em}}\qw & \gate{R^{(3)}}& \qw & \push{\rule{0em}{1.4em}}\qw & \gate{R^{(4)}} & \qw & \qw & \push{\rule{0em}{1.4em}}\qw & \push{\rule{0em}{1.4em}}\qw & \qswap & \qw & \qw & \qw \\
  &\lstick{2} &
  \qw & \qw & \qw & \ctrl{-1} & \gate{\mbox{ }H\mbox{ }} & \qw & \gate{R^{(2)}} & \qw & \qw & \gate{R^{(3)}} & \qw & \qw & \qw & \qw\qwx & \qw & \qswap & \qw \\
  &\lstick{1} &
  \qw & \qw & \qw & \qw & \qw & \ctrl{-2} & \ctrl{-1} & \gate{\mbox{ }H\mbox{ }}& \qw & \qw & \gate{R^{(2)}} & \qw & \qw & \qw\qwx &  \qw & \qswap\qwx &\qw \\
  &\lstick{0} & \qw & \qw & \qw & \qw & \qw & \qw & \qw & \qw &
  \ctrl{-3} & \ctrl{-2} & \ctrl{-1} & \gate{\mbox{ }H\mbox{ }} & \qw &
  \qswap \qwx & \qw & \qw &\qw \gategroup{2}{4}{5}{4}{.5em}{^\}}
  \gategroup{2}{5}{5}{5}{.5em}{^\}} \gategroup{2}{6}{5}{6}{.5em}{^\}}
  \gategroup{2}{7}{5}{7}{.5em}{^\}} \gategroup{2}{8}{3}{9}{.5em}{^\}}
  \gategroup{2}{10}{5}{10}{.5em}{^\}}
  \gategroup{2}{11}{4}{13}{.5em}{^\}}
  \gategroup{2}{14}{5}{14}{.5em}{^\}}\\
}
$$
\end{footnotesize}
\caption{A circuit for QFT over $n=4$ qubits. Note that the qubits are
  numbered from bottom to top, starting from $0$ up to $n-1$.}
\label{fig:circuitA}
\end{figure}
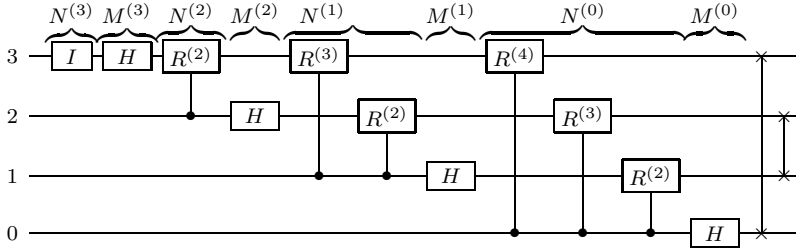

   In Fig.~\ref{fig:circuitA} we represent the QFT circuit over four
   qubits in terms of Hadamard, controlled gates $R^{(u)}$ and swap
   operations.  Note that matrices $M^{(s)}$ and $N^{(s)}$ are also
   shown on the top. An alternative presentation of the circuit may be
   obtained by interchanging some of the gates that do not involve
   operations on the same qubits.

   Now we address the computational complexity of
   Algorithm~\ref{alg:qft}. We assume that the initialization is done
   in negligible time---which is a reasonable assumption, since in
   most applications known so far the QFT is applied on a state of the
   computational basis, or on the output of some earlier step of some
   algorithm.  In the main part of the algorithm (the outer loop) we
   have $n(n+1)/2$ steps.  Therefore, this part of the algorithm is
   $O(n^2)$. The last part of the algorithm consists only on $O(n)$
   swap operations.  We conclude that the QFT has complexity $O(n^2)$
   or, equivalently, $O(\log^2 N)$, in terms of one and two qubit
   operations. It is quite simple to show that this complexity does
   not change when calculated in terms of universal gates. One just
   needs to recall that a controlled gate can be decomposed into two
   CNOTs and three one-qubit gates, and that a swap gate can be
   decomposed into three CNOTs.

   As an intermediate step before deducing the Approximate QFT, we may
   consider a classical algorithm for the Approximate FFT. This is
   quite similar to Algorithm~\ref{alg:copp}, the only difference
   being that instead of $\w^{(b_s \ldots b_{n-1} 0 \ldots 0)_2}$ in
   the second term on the right hand side of Eq.~(\ref{eq:alg}), we
   have $\w^{(b_s \ldots b_{\operatorname{min}(s+m-1,n-1)} 0 \ldots
     0)_2}$. By repeating all the process of finding the generic
   matrices $P^{(s)}$ and decomposing them, we find out that this
   difference reflects only on Eq.~\eqref{eq:NProd}---in the
   approximate algorithm the productory ranges from $t=s+1$ to
   $\operatorname{min}(s+m-1,n-1)$. Analogously to the exact
   algorithm, we may check that the Approximate QFT has complexity
   $O(mn)$ or, equivalently, $O(m\log N)$. In fact, the approximate
   version of QFT is not only simpler and faster, but also leads to
   more precise results in the presence of decoherence than its exact
   counterpart~\cite{BEST96a}.

\section{Conclusions}\label{sec:conclusions}

The QFT algorithm represents an important improvement in the
complexity of the classical algorithm, from $O(N\log N)$ to $O(\log^2
N)$. The acceleration provided by Approximate QFT algorithm goes even
further, as $O(m\log N)$, where $m$ is a parameter that can be taken
as $O(\log\log N)$. A practical difference between the classical and
the quantum FFT is that the latter provides the result of the
calculation as a superposition of quantum states, which cannot be
directly read according to the postulates of quantum
mechanics. However, a remarkable example showing the advantage of the
quantum version of FFT is the quantum algorithm for factorization of
large integers~\cite{Sho94a}. In this algorithm, the quantum FFT
subroutine plays an essential role on the \textit{exponential
  speed-up} over the best classical algorithms for integer
factorization.

In this paper, the building process of the QFT was exposed in detail,
generalizing Coppersmith's work. We started from the description of
the classical FFT and then obtained \textit{generic} unitary matrices
for each step of the algorithm. In order to get matrices simple enough
to represent feasible quantum operations, those generic matrices were
factored according to QR decomposition and the formulation in terms of
one- and two-qubit gates was obtained. The complexity of this
particular algorithm does not change when expressed in terms of
universal gates.

We argued that the Approximate QFT is also reobtained by the method
proposed here, with analogous calculations. Depending on the chosen
parameter, the number of matrices generated by the decomposition of
the approximate algorithm can be considerably lower than that of the
exact one. This reflects on the complexity of the quantum algorithm.

The deduction of the algorithms addressed in this paper is different
from previous works, and we hope it may provide insight for the
development of new efficient quantum algorithms. As future directions,
we are interested in checking whether other classical algorithms can
be analysed according to this technique, starting from its matrix form
and then decomposing it until simpler unitary matrices are obtained.

\paragraph*{Acknowledgments.}
The authors thank J.S.E.~Ortiz for helpful
discussions. F.L.M. acknowledges financial support from CNPq, as well
as previous financial support from FAPERJ, when this project was
initiated. R.P. thanks CNPq for financial support, grant
n. 306024/2008. F.D.S. is thankful to LNCC for the hospitality
during his stay. He also gratefully acknowledges the financial support
of UDESC.


\bibliographystyle{elsart-num} \bibliography{qft}

\end{document}